\documentclass[twocolumn,prl]{revtex4}

\ifx\pdfoutput\undefined
  \usepackage[dvips]{graphicx}
\else
  \usepackage[pdftex]{graphicx}
\fi

\usepackage{dcolumn}
\usepackage{amsmath}
\usepackage{latexsym}

\begin{document}

\title[Short Title]{Dynamics of Josephson junctions and single-flux-quantum
networks with superconductor-insulator-normal metal junction shunts}
\author{A.\,B.~Zorin}
\author{E.\,M.~Tolkacheva}
\author{M.\,I.~Khabipov}
\author{F.-I.~Buchholz}
\author{J.~Niemeyer}

\affiliation{Physikalisch-Technische Bundesanstalt, Bundesallee
100, 38116 Braunschweig, Germany}%

\date{December 23, 2005}

\begin{abstract}
Within the framework of the microscopic model of tunneling, we modelled the
behavior of the Josephson junction shunted by the
Superconductor-Insulator-Normal metal (SIN) tunnel junction. We found that the
electromagnetic impedance of the SIN junction yields both the
frequency-dependent damping and dynamic reactance which leads to an increase in
the effective capacitance of the circuit. We calculated the dc $I\textrm{-}V$
curves and transient characteristics of these circuits and explained their
quantitative differences to the curves obtained within the resistively shunted
junction model. The correct operation of the basic single-flux-quanta circuits
with such SIN-shunted junctions, i.e. the Josephson transmission line and the
toggle flip-flop, have also been modelled.


\verb  PACS numbers: 74.50.+r, 84.30.-r, 73.40.Gk
\end{abstract}
\maketitle

\section{Introduction}

Recently, the Josephson digital circuits operating on Single Flux Quantum (SFQ)
pulses \cite{Likh-Sem,Bunyk} have been considered as electronic circuits
suitable for integrating with Josephson qubits (see, for example,
Refs.\,\cite{Sem-Aver,Ohki}). Applying Josephson SFQ electronics for the
control and readout of the qubits has many advantages, including a high speed
of operation and low operating temperature in combination with rather small
dissipating power, allowing a sufficiently close location of the elements on
the chip. Moreover, this Rapid SFQ (RSFQ) electronics makes it possible to
process the input and output signals of the qubits directly on chip. This
possibility can extend the class of algorithms to be realized in the Josephson
quantum computer. Finally, a similar Josephson junction fabrication technology
for both, the RSFQ circuits and the qubits, is an essential prerequisite to
reach a joint circuit architecture.

The operating principle of the RSFQ circuits is based on the single
$2\pi$-leaps of the Josephson phase $\varphi$ in the overdamped junctions. Due
to large damping, the driven Josephson junctions never switch completely into
the phase-running regime with a large average voltage across the junction, but
generate short SFQ pulses, $V(t)=(\Phi_0/2\pi)d\varphi/dt$, with quantized
area, $\int V(t)dt=\Phi_0$, where $\Phi_0=h/2e\approx 2.07\times 10^{-15}$\,Wb
is the flux quantum \cite{Likh-Sem}. For the most manufacturable
Superconductor-Insulator-Superconductor (SIS) tunnel junctions with a not very
large critical current density $j_c$, operating at a temperature which is
notably below the critical temperature of the superconductor, $(T \lesssim 0.5
\,T_c)$, intrinsic damping is very small. Sufficient damping is therefore
achieved due to an external low-ohmic resistance $R_s$ shunting the junction,
so the McCumber-Stewart parameter \cite{McCum,Stew},
\begin{equation} \label{beta_c}\beta_c = \omega_c^2/\omega_p^2 =(2\pi /\Phi_0)I_c R_s^2 C,
\end{equation}
where $I_c$ is the critical current and $C$ is the junction capacitance, is
small, i.e. $\beta_c\lesssim 2$. At such values, the plasma resonance frequency
of the Josephson junction, $\omega_p = (2\pi I_c/\Phi_0 C)^{1/2}$, is just
slightly lower than the Josephson characteristic frequency $\omega_c =
(2\pi/\Phi_0)I_c R_s$ and the plasma oscillations are strongly damped.

Contrary to this, the operation of Josephson qubits requires vanishing damping
in both the qubit's junctions and in the environment, including the coupled
control and readout circuits. The effect of unsuppressed damping is, however,
dramatic and appears in fast decoherence of the qubit. Especially, the noise
resulting from damping at frequencies around the characteristic frequency of
the qubit, $\Omega = (E_0-E_1)/\hbar$ (typically, about 10\,GHz), causes
intensive relaxation, while the low-frequency components of the noise lead to a
dephasing of the qubit (see, e.g. the review \cite{Makhlin}). At short
decoherence times, qubit manipulation is unfeasible. This situation can to a
certain degree be softened by weakening and/or switching on and off the
coupling \cite{Clarke-switch} and by operating the qubit in the optimal points
where the qubit is immune to the external noise in the linear order
\cite{Vion,Z-JETP}. In the case of the RSFQ circuit interface coupled to the
qubit, the low resistance of the junction shunts $R$ is the source of large
broad-band current noise ($\propto R^{-1}$) acting on the qubit. Moreover, this
noise is generated by the resistors even in the quiescent (zero-voltage) state
of the Josephson junctions. So, the problem of reducing the noise of RSFQ
circuits coupled to the qubit has to be solved radically.

\begin{figure}[b]
\begin{center}
\includegraphics[width=3.0in]{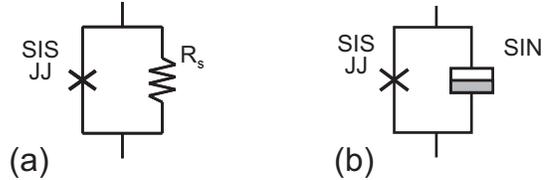}
\caption{Electrical circuit diagram of (a) an resistively shunted and (b)
SIN-shunted SIS Josephson junction (JJ).} \label{EqvSchm}
\end{center}
\end{figure}

A promising approach based on the possibility of frequency-dependent damping of
SIS junctions by means of their shunting with Superconductor-Insulator-Normal
metal (SIN) tunnel elements (see Fig.\,1) has recently been proposed in
Ref.\,\cite{Z-APL}. In particular, it was shown that shunting by a non-linear
resistance drastically improves the noise characteristics of the circuit and
ensure sufficient damping to achieve an almost non-hysteretic shape of the
$I$-$V$ curves. The applied model was, however, rather simplified and the
special aspects of dynamical processes not investigated. In this paper, we
address the problem of a microscopic model describing the SIS+SIN circuit, as
well as the analysis of the dynamics of the basic RSFQ networks within the
scope of this model. The aim of this work was to demonstrate the full
functionality and applicability of these circuits for the Josephson qubits.

\section{Simple model}

Due to the very large values of the zero-bias differential resistance and the
strongly nonlinear dc $I$-$V$ characteristic of the SIN junctions at low
temperature ($T\ll T_c$), their noise is small in the range of frequency up to
$\omega_g \equiv \Delta/\hbar$, where $\Delta$ is the energy gap of the
superconducting electrode of the SIN junction. If the qubit's characteristic
frequency is sufficiently small, $\Omega < \omega_g$, the influence of this
noise on the qubit is expected to be weak. On the other hand, if $\omega_g$ is
lower than the characteristic Josephson frequency $\omega_c = (2\pi/\Phi_0)I_c
R$, where $R$ is the asymptotic resistance of the SIN junction at large bias
($V\gg V_g \equiv \Delta/e$), a large effective damping of the Josephson
junction is still possible at sufficiently small values of parameter $\beta_c$
given by Eq.\,(\ref{beta_c}) with replacement $R_s \rightarrow R$ \cite{Z-APL}.
The capacitance $C$ in this equation is equal to the sum of the capacitances of
the SIS and SIN elements, $C_{\textrm{SIS}}+C_{\textrm{SIN}}$.

For an analysis of the $I$-$V$ characteristics of the SIN-shunted Josephson
junctions the simplified equation of motion with nonlinear conductance term was
numerically solved in Ref.\,\cite{Z-APL}, viz.,
\begin{equation} \label{motion}
C\frac{dV}{dt} + I_{\textrm{SIN}}^{\textrm{dc}}(V) + I_c\sin \varphi= I,
\end{equation}
where the instant voltage
\begin{equation} \label{motion-add}
V(t)=\frac{\Phi_0}{2\pi}\frac{d\varphi}{dt}.
\end{equation}
Here, $I_{\textrm{SIN}}^{\textrm{dc}}(V)$ is the dc $I$-$V$ characteristic of
the SIN junction with the instant voltage $V(t)$ as an argument.
Equation\,(\ref{motion}) is a modified Resistively Shunted Junction (RSJ) model
equation \cite{McCum,Stew}, relating the terms taken at a given instant.
Earlier, a similar model was applied for describing the quasiparticle damping
in SIS junctions by Prober et al. \cite{Prober}, who approximated the nonlinear
conductance term by three straight line segments.

Equation\,(\ref{motion}) is, however, too rough and does not describe correctly
the processes as the decay of plasma oscillations caused by a short kick of the
phase in the zero-current-biased circuit. In fact, after fast exponential decay
and approaching the sufficiently low level ($\approx V_g$), the amplitude $V_A$
of the oscillations $V(t)=V_A(t) \cos \omega_p t$ starts to decrease much
slower, because of small damping being available for the small instant values
of voltage $V$ according to Eq.\,(\ref{motion}). On the other hand, at
sufficiently small $\beta_c \lesssim 1$, these oscillations should decay very
fast, independent of their amplitude. To avoid this discrepancy, a more
elaborate (microscopic) model of the circuit in Fig.\,1b has to be applied.

\section{Microscopic model}

We will derive the current through the SIN junction starting from the tunneling
Hamiltonian of Cohen, Falicov and Phillips \cite{Cohen},
\begin{equation} \label{H-main}
H = H_0 + H_T \equiv H_S + H_N + eVN_S+ H_T.
\end{equation}
Thereby, term $H_{S}$ ($H_{N}$) is the Hamiltonian of the superconductor
(normal) electrode; term $eVN_S$ includes the finite voltage $V$ across the
junction. The tunneling is described by the Hamiltonian
\begin{equation} \label{Ht}
H_T = \sum_{k,q} T_{k,q}c_{k}^{\dag} d_q +  T^*_{k,q}c_{k}d_q^{\dag}
\end{equation}
and is considered to be a small perturbation. Here $c_k$ and $c_{k}^{\dag}$
are, respectively, the destruction and creation operators for an electron in
state $k$ in the superconductor electrode, while $d_q$ and $d_q^{\dag}$ are the
corresponding operators for the normal electrode. The nonzero (generally,
non-constant) voltage $V$ leads to the additional ac phase factors for, say,
the superconductor-electrode operators, $c_k \rightarrow c_k e^{i\varphi(t)/2}$
and $c_{k}^{\dag} \rightarrow c_{k}^{\dag} e^{-i\varphi(t)/2}$, with
$\varphi(t) = (2\pi/\Phi_0)\int^tV(t')dt'$.

The number operators are equal to
\begin{equation} \label{NsNn}
N_S = \sum_{k} c_{k}^{\dag} c_k,\quad N_N = \sum_{q} d_{q}^{\dag} d_q,
\end{equation}
so the tunneling current is expressed via the expectation values of the
operators $\dot{N}_S$ and $\dot{N}_N$,
\begin{equation} \label{I-tunn}
I_{\textrm{SIN}} = \langle \hat{I} \rangle =  e \langle \dot{N}_S\rangle =
-e\langle \dot{N}_N\rangle.
\end{equation}
Applying the perturbation theory assuming adiabatic turning on from the past of
the interaction $H_T$ we obtain the standard first-order result (see, for
example, Ref.\,\cite{Rickayzen}),
\begin{equation} \label{I-tunn2}
I_{\textrm{SIN}} = -\frac{i}{\hbar}\int_{-\infty}^t e^{+0(t'- t)}
\langle [ \hat{I}(t),H_T(t')]\rangle_0 \,dt',
\end{equation}
where $\langle...\rangle_0$ denotes averaging over the ensemble $H_0$.
Introducing the retarded Green's function
\begin{eqnarray} \label{Kt1} &&K(t-t')=i(2e/\hbar^2)\theta(t-t') \exp[+0(t'-t)]\qquad \nonumber\\
&& \times \sum_{k,q,k',q'} T_{k,q}T^*_{k',q'} \langle[ c_{k}^{\dag}(t) d_q(t),
c_{k'}(t')d_{q'}(t')^{\dag} ]\rangle_0,
\end{eqnarray}
where $\theta(t)$ is the Heaviside step function, we finally arrive at the
convolution integral expression for the current
\begin{equation} \label{I-tunn3}
I_{\textrm{SIN}} = \int_{-\infty}^t K(t-t')\sin
\frac{\varphi(t)-\varphi(t')}{2} \,dt'
\end{equation}
with function $K$ playing the role of a memory kernel. One can see that the
tunneling current depends both on the instant value of phase $\varphi(t)$ and
on its values in the past, $\varphi(t')$, $t'\leq t$, so Eq.\,(\ref{I-tunn3})
describes the causal physical process.

Actually, the relations similar to Eqs.\,(\ref{Kt1}) and (\ref{I-tunn3}) yield
the quasiparticle component of the current in the microscopic theory of the
Josephson tunneling developed by Werthamer \cite{Wert} and Larkin and
Ovchinnikov \cite{LarOvch} and presented in the time domain by Harris
\cite{Harris3}. For our case of the SIN junction, the kernel $K(\tau)$ can be
easily found from its Fourier transform giving the well-known dc $I\textrm{-}V$
curve of the junction. In fact, in the special case of a constant voltage bias
$V\equiv V_0$ the phase $\varphi$ runs linearly, i.e. $\varphi=\omega_v t +
\textrm{const}$, where $\omega_v= (2\pi/\Phi_0)V_0$, and Eq.\,(\ref{I-tunn3})
yields the dc current as a Fourier integral,
\begin{equation} \label{I-tunn4}
I_{\textrm{SIN}}^{\textrm{dc}}(V)=-I_{\textrm{SIN}}^{\textrm{dc}}(-V)=
\int_0^{\infty} K(\tau)\sin(\omega_v \tau/2)\,d\tau.
\end{equation}
The BCS-theory-based expression for $I_{\textrm{SIN}}^{\textrm{dc}}$ is given
by the integral over the states in the energy representation,
\begin{equation} \label{I-tunn5}
I_{\textrm{SIN}}^{\textrm{dc}}(V)= \frac{1}{eR}\int_{-\infty}^{\infty}
\frac{|E|[f(E)-f(E+eV)]}{(E^2-\Delta^2)^{1/2}}\,dE,
\end{equation}
where the actual range of integration is $|E|> \Delta$ and where
$f(E)=(1+\exp(E/k_BT))^{-1}$ is the Fermi function (see, e.g.,
Ref.\,\cite{Tinkham}). Applying the reverse Fourier transform to
Eq.\,(\ref{I-tunn4}) we obtain
\begin{equation} \label{K-rev-F}
K(\tau) = \frac{2}{\pi}\int_0^{\infty} I_{\textrm{SIN}}^{\textrm{dc}}(\Phi_0
\omega/\pi)\sin\omega \tau\,d\tau, \quad \tau\geq 0.
\end{equation}
In the case of zero temperature, $T=0$, when the $I\textrm{-}V$ curve has the
hyperbolic shape,
\begin{equation} \label{IVC1}
I_{\textrm{SIN}}^{\textrm{dc}}(V)=[2\theta(V)-1]
\theta(|V|-V_g)(V^2-V^2_g)^{1/2}R^{-1},
\end{equation}
the integral in Eq.\,(\ref{K-rev-F}) can be computed explicitly yielding
\begin{equation} \label{K-t}
K(\tau) = - (2\hbar/eR)\delta'(\tau)+K_1(\tau),
\end{equation}
where
\begin{equation} \label{K-t2}
K_1(\tau)= -(\Delta/eR)\theta(\tau)J_1(\omega_g \tau)/\tau.
\end{equation}
Here $\delta'$ is the time derivative of the Dirac delta-function and $J_1$ is
the Bessel function of the first order. The first term in Eq.\,(\ref{K-t})
describes the linear component of damping due to the ohmic asymptotic at high
voltage, $|V|\gg V_g$, while the second term, given by Eq.\,(\ref{K-t2}),
describes the dispersive damping due to a strong nonlinearity of the
$I\textrm{-}V$ characteristic in the vicinity of $V_g$. (Compare the shape of
the kernel Eq.\,(\ref{K-t2}) with $(\pi \Delta^2/\hbar e R)
\theta(\tau)J_1(\omega_g \tau)Y_1(\omega_g \tau)$, where $Y_1$ is the Bessel
function of the second kind, obtained by Harris \cite{Harris3} for the SIS
junction.) In the case of nonzero temperature, $T\neq 0$, both integrals in
Eqs.\,(\ref{I-tunn5}) and (\ref{K-rev-F}) yielding the kernel $K(\tau)$ should
be calculated numerically.

\begin{figure}[t]
\begin{center}
\includegraphics[width = 3.0in]{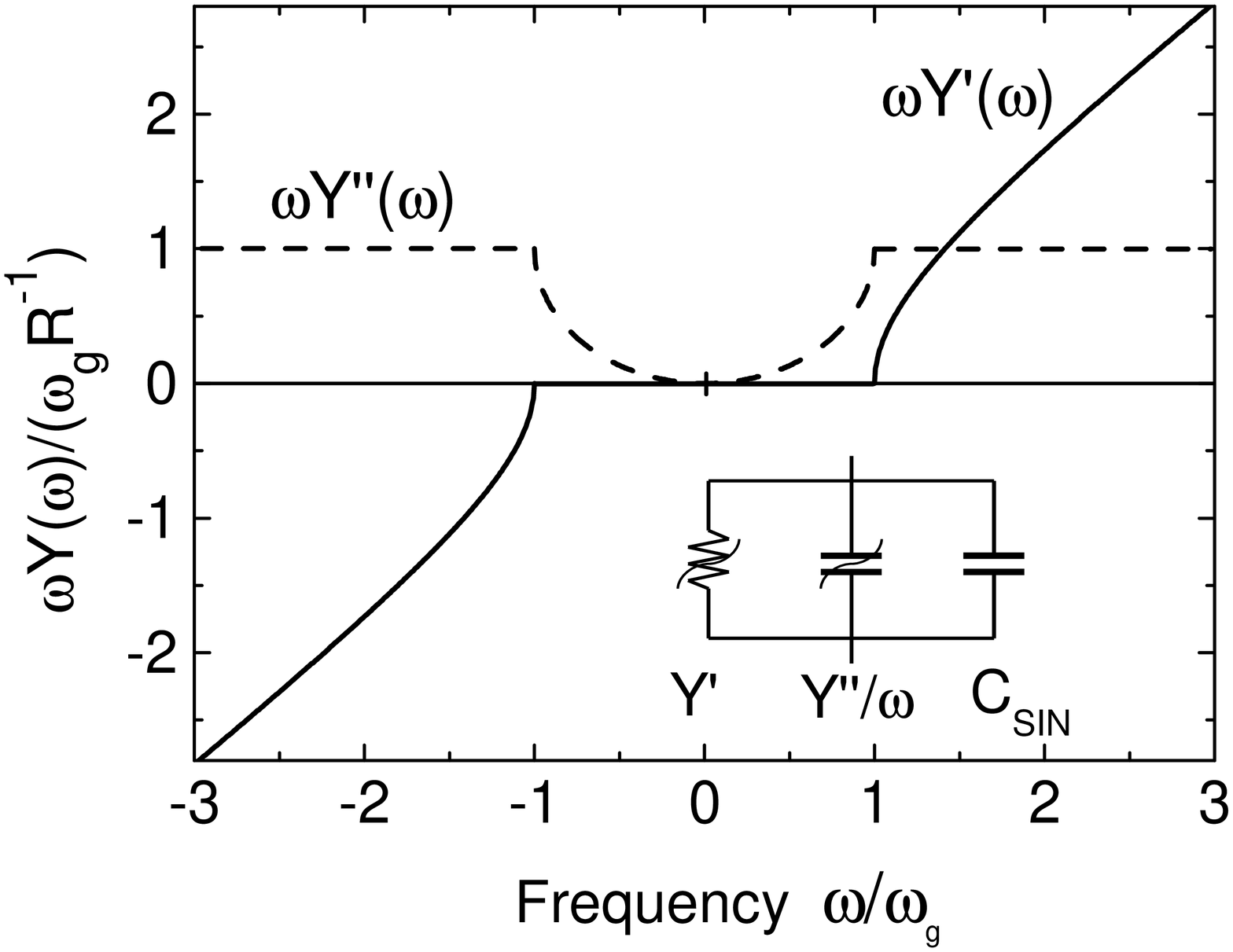}
\caption{Real (solid line) and imaginary (dashed line) parts of the product
$\omega Y(\omega)= \omega [Y'(\omega)+iY''(\omega)]$ calculated for the case of
zero temperature, $T=0$. The solid line curve also shows the shape of the dc
$I$-$V$ curve of the SIN junction (see the first equality in Eq.\,(\ref{Y1})).
The increase in temperature $T$ leads to a rounding of the sharp corners of
these curves at $\omega = \pm\omega_g$. The inset shows the equivalent
electrical circuit of the SIN junction, which comprises (from left to right)
the dynamic bias-dependent conductance, the dynamic capacitance (both frequency
dependent) and the geometrical capacitance of the junction barrier.} \label{Y}
\end{center}
\end{figure}

The sine Fourier transform of the kernel $K(\tau)$
Eq.\,(\ref{I-tunn4}) and the corresponding cosine transform give
the real and imaginary parts of the junction's complex
admittance, respectively, $Y = Y' + iY''$, viz.,
\begin{equation} \label{Y1}
\omega Y'(\omega) = \frac{e}{\hbar} I_{\textrm{SIN}}^{\textrm{dc}}(\hbar
\omega/2e)= \frac{e}{\hbar}\int_0^{\infty} K(\tau)\sin\omega \tau\,d\tau,
\end{equation}
\begin{equation} \label{Y2}
\omega Y''(\omega) =\frac{e}{\hbar} \int_0^{\infty}
K(\tau)(\cos\omega \tau\ -1)\,d\tau.
\end{equation}
These relations arise from Eq.\,(\ref{I-tunn3}) on the assumption of a small ac
voltage $V = v_a \cos \omega t$ yielding $\varphi= a \sin \omega t$ with
$a=2\pi v_a/\omega \Phi_0\ll 1$. The corresponding real and imaginary parts of
the admittance Eqs.\,(\ref{Y1}) and (\ref{Y2}) determined by the casual kernel
$K(\tau)$ obey the Kramers-Kronig relations.

As expected, the nonlinear dc $I\textrm{-}V$ curve
$I_{\textrm{SIN}}^{\textrm{dc}}(V)$ yields according to Eq.\,(\ref{Y1}) the
frequency-dependent damping $Y'(\omega)$. The odd function $Y''(\omega)$
describing the junction's reactance is positive for $\omega>0$, so the SIN
tunnel junction behaves as an equivalent frequency-dependent dynamic
capacitance $\tilde{C}(\omega) = Y''(\omega)/\omega$ with the
frequency-dependent losses $Y'(\omega)$, which is added up with the geometrical
capacitance of the sandwich $C_{\textrm{SIN}}$. For the case of zero
temperature, $T=0$, the plots of the components Eqs.\,(\ref{Y1}) and (\ref{Y2})
are presented in Fig.\,2, where the inset shows the equivalent electrical
circuit of the SIN junction. Note that at $\omega \rightarrow 0$, the value of
the dynamic capacitance is finite, i.e., $\tilde{C} = (2\omega_g R)^{-1}$. For
rather transparent barriers (for example, for the SIN junctions with an Al
superconducting electrode and a specific resistance of barrier $\rho =
30\,\Omega\cdot \mu\textrm{m}^2$) this capacitance is comparable with the
geometrical capacitance of the barrier (about 50\,fF/$\mu\textrm{m}^2$).

\section{Equation of motion and its solving}

For computing the SIS junction current in the circuit Fig.\,1b we can naturally
apply either the microscopic tunnel model \cite{Wert,LarOvch} in the form of
\cite{Harris3} or the simpler adiabatic model of the Josephson junction giving
$I_{\textrm{SIS}}=I_c \sin \varphi$ \cite{Jos}. In our case of significant
total damping due to sufficiently low resistance $R$ of the SIN junction
\cite{Z-APL}, the quasiparticle current of the SIS junction can be neglected.
Moreover, since the characteristic frequency $\omega_c$ is appreciably lower
than the gap frequency of the SIS junction $\omega_g^{\textrm{SIS}}\equiv
2\Delta_{\textrm{SIS}}/\hbar$ ($\Delta_{\textrm{SIS}}$ is the energy gap of the
SIS junction electrodes, which in the most favorable case should be much larger
than $\Delta$ \cite{comment}), one can neglect the dispersion of the
supercurrent. This dispersion is essential at frequencies $\omega \approx
\omega_g^{\textrm{SIS}}$ and manifests itself as the logarithmic Riedel peak
\cite{Riedel}. Therefore, the model combining the microscopic description of
the SIN junction and the adiabatic description of the SIS junction (with
constant amplitude of supercurrent $I_c$) is adequate for the circuit Fig.\,1b.
So, the equation of motion takes the form
\begin{eqnarray} \label{motion2}
&&C\left(\frac{\Phi_0}{2\pi}\right)\frac{d^2\varphi}{dt^2}
+\frac{1}{R}\left(\frac{\Phi_0}{2\pi}\right)\frac{d\varphi}{dt}+  I_c\sin
\varphi
\nonumber\\
&& + \int_{-\infty}^t K_1(t-t')\sin
\frac{\varphi(t)-\varphi(t')}{2} \,dt' = I.
\end{eqnarray}
Here we decomposed the kernel $K(\tau)$ into two parts in accordance with
Eq.\,(\ref{K-t}), presenting the asymptotic constant contribution of damping
separately, by the second term on the left-hand side.

\begin{figure}[t]
\begin{center}
\includegraphics[width = 3.0in]{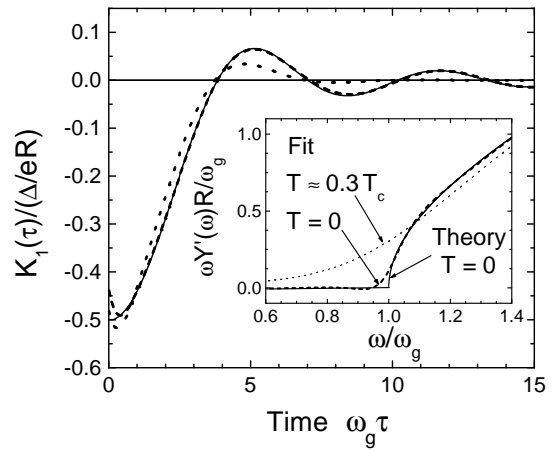}
\caption{Fitting of kernel $K_1(\tau)$ Eq.\,(\ref{K-t2}) (solid line) by the
Dirichlet series Eq.\,(\ref{K-Dir}) with three terms (dashed line). The
inessential discrepancy of the curves at small values of time $\tau$ is related
to the imperfection of the fit of the Fourier transform Eq.\,(\ref{Y1}) at high
frequencies, $\omega\gg\omega_g$. Fitting of this Fourier transform, i.e. the
dc $I$-$V$ curve Eq.\,(\ref{I-tunn4}), in the most critical range of
frequencies $\omega \sim \omega_g$ is shown in the inset. The dotted line
curves in both plots show the shape of the kernel and the $I$-$V$ curve given
by two terms in the series Eq.\,(\ref{K-Dir}) (coefficient $B_2 = 0$) that
roughly corresponds to the case of a finite temperature of $T \approx
0.3\,T_c$.} \label{fit}
\end{center}
\end{figure}

Generally, the solving of an integro-differential equation with a slowly
decaying kernel (see Eq.\,(\ref{K-t2})) is a difficult task, because at each
time step $t\rightarrow t +\Delta t$ of the numerical integration one has to
take a convolution integral over the time interval from $-\infty$ to $(t
+\Delta t)$, that dramatically slows down such calculations. However, the
calculations can be significantly accelerated if one approximates the kernel by
a finite Dirichlet series \cite{Apostol}, because the exponential shape of the
kernel makes it possible to avoid the time-consuming direct integration at each
step. Instead, only small corrections to the convolution integrals are computed
at each time step. This procedure was realized by Odintsov et al.
\cite{Odintsov} for the kernels derived in the microscopic model of the SIS
junction \cite{Wert,LarOvch}. For our case of an SIN junction having frequency
dispersion of a relatively simple shape (see Fig.\,2), the Dirichlet series can
contain only a few terms, but still describe the dynamics adequately,
\begin{equation} \label{K-Dir}
K_1(\tau)= \frac{\Delta}{eR} \,\textrm{Re} \sum_{n=1}^{N}B_n
e^{p_n \tau},\quad \tau \geq 0;
\end{equation}
here coefficients $B_n$ and $p_n$ are the complex numbers with Re$(p_n)<0$.

Figure\,3 shows the result of the kernel fitting by the Dirichlet series
Eq.\,(\ref{K-Dir}) with $N=3$ for the set of frequencies:
Im$p_1=0.95\,\omega_g$, Im$p_2=\omega_g$ and Im$p_3=1.05\,\omega_g$. One can
see that the approximating function captures well both the behavior of kernel
$K_1(\tau)$ and its Fourier transform (shown in the inset). Interestingly, the
two-term approximation of the series Eq.\,(\ref{K-Dir}) with the frequencies
Im$p_1=0.95\,\omega_g$ and Im$p_3=1.05\,\omega_g$ (shown by dotted lines)
yields a reasonable approximation of the dc $I$-$V$ characteristic for finite
temperature. The obtained set of coefficients $B_n$ and the damping factors
Re$p_n$ in Eq.\,(\ref{K-Dir}) made it possible to reduce the numerical solving
of the integro-differential equation Eq.\,(\ref{motion2}) to almost that of an
ordinary differential equation. For the single junction circuits the
simulations were technically performed by applying the fourth-order Runge-Kutta
method. For the multi-junction circuits the simulations were done with the help
of the fitted program code PSCAN developed earlier for both the RSJ and the
tunnel junction models \cite{Odintsov,PSCAN}.

\section{Josephson junction networks with SIN shunts}

\begin{figure}[t]
\begin{center}
\includegraphics[width = 3.1in]{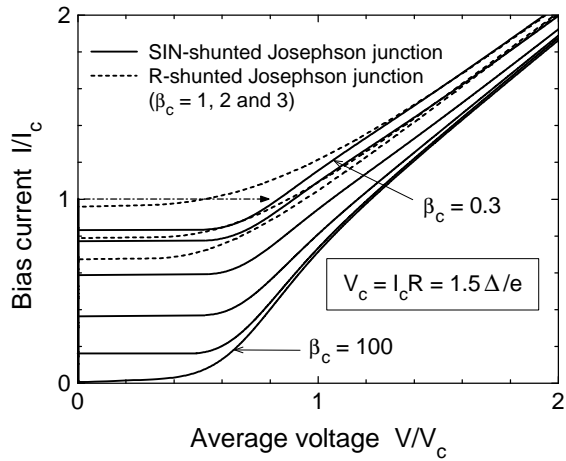}
\caption{The autonomous $I$-$V$ curves calculated within the frame of the
microscopic SIS+SIN model and the RSJ model. The values of the McCumber-Stewart
parameter $\beta_c$ for the SIN-shunted junction are 0.3, 0.5, 1, 2, 5 and 100
(solid lines), while for the RSJ model the values are 1, 2 and 3 (dashed
lines), in sequence from the top curve to the bottom curve. The dash-dot arrow
indicates the switching from the superconducting to the resistive state due to
the current bias regime. The bottom curve calculated for large $\beta_c = 100
\gg 1$ practically coincides with the $I\textrm{-}V$ curve of the stand-alone
SIN junction. The fragment of this curve is also shown in the inset of Fig.\,3
by dotted line. The value of the characteristic voltage is equal to
$V_c=1.5\Delta/e$ in this plot as well as in the plots which follow.}
\label{VAX}
\end{center}
\end{figure}

Applying the described procedure of solving the equation of motion, we first
found the dc $I$-$V$ characteristics of the circuit Fig.\,1b biased by a
constant current. The resulting curves are shown in Fig.\,4, where they are
compared with the $I\textrm{-}V$ curves given by the RSJ model (Fig.\,1a). One
can see that the shapes of these curves are qualitatively similar, although the
values of $\beta_c \approx 0.3-0.5$ ensuring sufficiently small hysteresis in
the curves for the SIN-shunted junction are appreciably smaller than the
corresponding values in the RSJ model ($\beta_c \approx 1-2$). Moreover, the
former curves exhibit characteristic plateaus at $V \lesssim \Delta/e$.
Interestingly, the size of such plateaus developed at small $\beta_c$ is
somewhat smaller than that obtained in the simplified (phenomenological) model
of the SIN junction (cf. Fig.\,1 of Ref.\,\cite{Z-APL}).

\begin{figure}[t]
\begin{center}
\includegraphics[width = 3.10in]{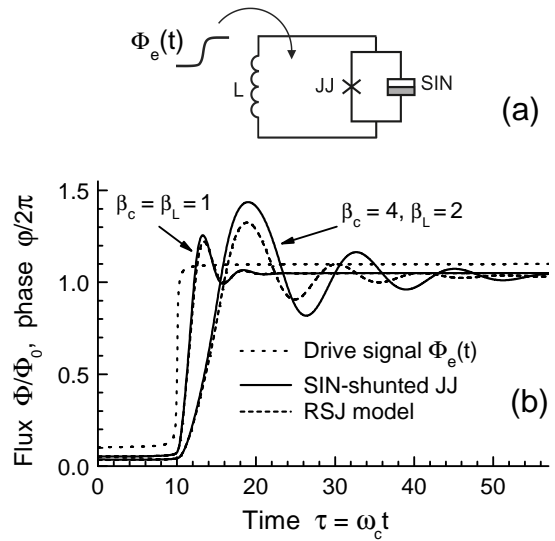}
\caption{(a) Electric diagram of the SIN-shunted Josephson junction inserted in
the superconducting loop with the dimensionless inductance $\beta_L=2$ and
driven by a step-pulse of the magnetic flux. (b) Switching characteristics of
the SIN-shunted (solid lines) and resistively shunted (dashed lines) Josephson
junctions induced by the short pulse (dotted line) of flux of unit amplitude
($\Phi_e(t)/\Phi_0 = 0.1 \textrm{ at } \omega_c t < 10 \textrm{ and } 1.1
\textrm{ at } \omega_c t > 10$) for two sets of values of $\beta_c$ and
$\beta_L$. The kernel $K_1$ is approximated by two terms in
Eq.\,(\ref{K-Dir}).} \label{switching}
\end{center}
\end{figure}

The dynamical process of the jump of the Josephson phase by $2\pi$ was modelled
in the circuit shown in Fig.\,5a. The inductance of the superconducting loop
$L$ closing the shunted SIS junction is comparable with the Josephson
inductance, i.e. the dimensionless parameter
\begin{equation} \label{betaL}
\beta_L= 2\pi I_c L/\Phi_0
\end{equation}
is equal to $1-2$. A short step-pulse of magnetic flux of the magnitude of
$\Phi_0$ was applied to the loop. The transient behavior caused by this pulse
is shown in Fig.\,5b. One can see that for the case of sufficiently large
damping, $\beta_c=\beta_L=1$, the curves for the microscopic SIN and
phenomenological RSJ models practically coincide. For smaller damping, i.e. for
the values $\beta_c=4,\: \beta_L=2$ giving the bare resonance frequency of the
circuit Fig.\,1a equal to $\omega_0=\omega_c
[\beta_c^{-1}(1+\beta_L^{-1})]^{1/2} \approx 0.61\,\omega_c \approx
0.92\,\omega_g < \omega_g$, the difference between the corresponding curves is
substantial. The transient behavior of the SIN-shunted junction shows
oscillations of larger amplitude and smaller frequency, compared to the RSJ
model. The reason for this behavior is the effect of the dynamic capacitance
$\tilde{C}(\omega)$ of the SIN junction (see Fig.\,2), giving $\omega_c
R\tilde{C}\approx 1.4$. This leads to a 20\% decrease in the resonance
frequency $\omega_0$. The corresponding effective value of the McCumber-Stewart
parameter $\beta_c$ becomes therefore respectively larger.

We also present the results of simulations of the basic RSFQ circuits, i.e. the
Josephson transmission line (JTL) and the toggle flip-flop (TFF)
\cite{Likh-Sem}. Figure\,6a shows the electric circuit diagram of the JTL
consisting of the chain of SIN-shunted Josephson junctions connected in
parallel by relatively small superconducting inductances $L$. An SFQ step pulse
of flux is applied to the leftmost loop of the line and causes sequential
triggering of $2\pi$-leaps in the junctions $J1$, $J2$, $J3$ and $J4$. As a
result, the SFQ voltage pulse is transferred along the line with a small time
delay, $\sim 2\pi/\omega_c$, on each cell (see Fig.\,6b).

\begin{figure}[t]
\begin{center}
\includegraphics[width = 3.2in]{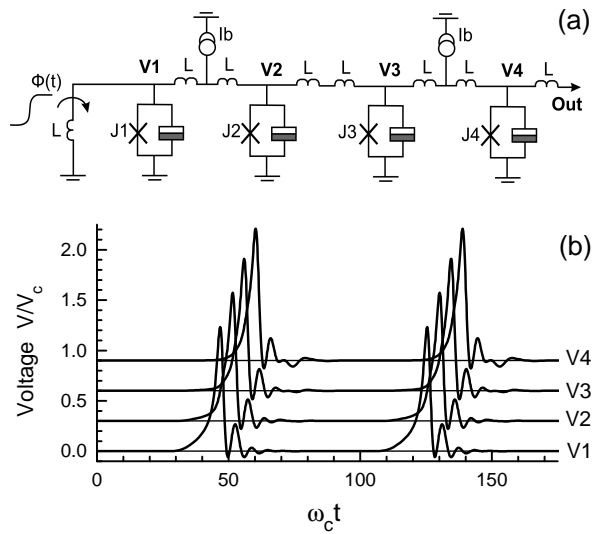}
\caption{(a) Electric circuit diagram of the 4-cell Josephson transmission line
constructed from SIN-shunted Josephson junctions with identical critical
currents $I_c$. The McCumber-Stewart parameter of these junctions is equal to
$\beta_c=0.3$. The bias currents $Ib$ are equal to $0.9\,I_c$. The value of
each inductance is $L=0.4\, \Phi_0/(2\pi I_c)$. (b) The SFQ voltage pulses on
the JTL cells, corresponding to the propagation of the $2\pi$-leap of the
phase; the voltages across different cells are offset for clarity.}
\label{JTL-line}
\end{center}
\end{figure}

Figure\,7a shows the electric circuit diagram of the TFF. Due to the
appreciable value of the storing inductance $L3$, this circuit has two stable
states that differ by the direction of the dc current circulating in the loop
of the interferometer $J1$-$L3$-$J4$. The SFQ pulses arriving at the input port
cause an alternating switching of the TFF, while the auxiliary junctions $J2$
and $J3$ prevent a back-reaction of the interferometer on the SFQ pulse source
\cite{Likh-Sem}. The switching of the direction of the current circulating in
the interferometer loop leads to a rectangular pulse of flux induced in
inductance $L3$ (see Fig.7\,b). Such pulse, as a control signal, can, for
example, be applied to the loops of the Josephson flux qubit of the double
SQUID configuration \cite{Chiarello}.

\begin{figure}[t]
\begin{center}
\includegraphics[width = 3.0in]{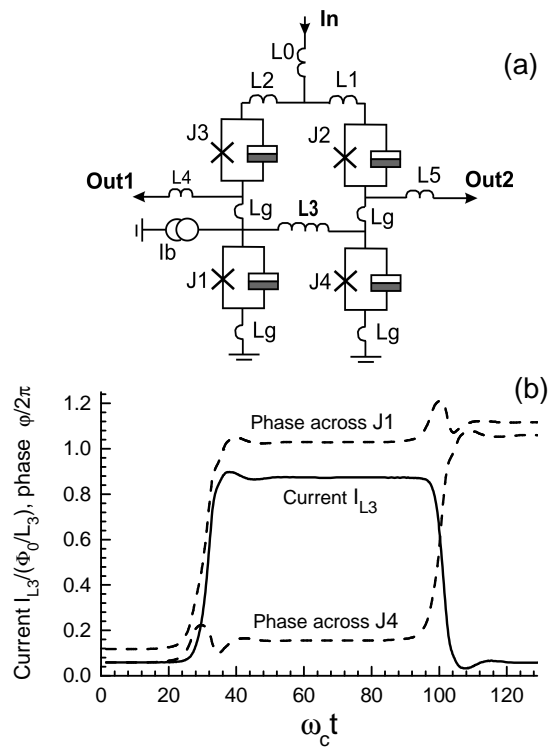}
\caption{(a) Electric circuit diagram of TFF constructed from the SIN-shunted
Josephson junctions $J1$, $J2$, $J3$ and $J4$ having the critical currents
$I_{c1}=1.35\,I_{c0}$, $I_{c2}=1.17\,I_{c0}$, $I_{c3}=1.55\,I_{c0}$ and
$I_{c4}=1.4\,I_{c0}$, respectively. The values of inductances $L0$, $L1$, $L2$,
$L3$, $L4$, $L5$ and $Lg$ expressed in units $L_u=\Phi_0/(2\pi I_{c0})$ are
0.4, 0.2, 0.2, 3.0, 0.8, 0.8, and 0.05, respectively. The bias current $Ib$ is
equal to $1.31\,I_{c0}$. (b) The time dependence of the current flowing through
the storing inductance $L3$ (solid line) and the $2\pi$-leaps of the phases on
the junctions $J1$ and $J4$ (dashed lines). Here the frequency
$\omega_c=(2\pi/\Phi_0)I_{c0}R$, where the tunnel resistance $R$ is similar for
all SIN junctions.} \label{JTL-line}
\end{center}
\end{figure}

\section{Discussion}

In summary, we have developed the microscopic model of the Josephson junction
shunted by the nonlinear element based on the SIN tunnel junction. The behavior
of the circuit is adequately described by the integro-differential equation
with a memory kernel reflecting the casual dependence of the tunneling current
on the phase across the junction. In contrast to the simplified model proposed
in Ref.\,\cite{Z-APL}, this model captures the expected features in the
behavior of the SIN-shunted Josephson circuits, including a clear dependence of
the decay of plasma oscillations on their frequency.

Besides the frequency-dependent damping, the SIN-junction shunts have the
effect of an enhanced capacitance. The resulting capacitance includes both the
geometrical capacitance of the junction and the dynamic frequency-dependent
capacitance that affects the shape of the $I\textrm{-}V$ curve increasing the
hysteresis. That is why the calculated $I\textrm{-}V$ curves are almost similar
to those given by the RSJ model for somewhat larger values of the
McCumber-Stewart parameter, i.e. $\beta_c^{\textrm{RSJ}} = 1\textrm{-}3$.
Still, the values $\beta_c \approx 0.3$ ensuring a sufficiently small
hysteresis in SIN-shunted Josephson junctions and the functionality of the
multi-junction circuits seem to be feasible. The most challenging condition to
be met is the realization of high-quality SIN junctions with a high
transparency of the tunnel barrier, having a specific resistance $\rho \lesssim
30\,\Omega\cdot \mu$m$^2$ \cite{Z-APL}. The recent experiments with
Nb-AlO$_\textrm{x}$-Al (at $T \geq 1.4$\,K) and Al-AlO$_\textrm{x}$-Cu (at $T
\leq 1$\,K) junctions have shown that the ratio of the zero-bias resistance to
the asymptotic tunnel resistance can achieve sufficiently large values, i.e.
$>\,30-50$ \cite{Balashov,Balashov2}. (This behavior of the SIN junctions can
be roughly described by the two-term approximation of the memory kernel
Eq.\,(\ref{K-Dir}).) Using such SIN junctions as shunts for Nb SIS junctions
can make it possible to significantly reduce the noise of the circuits in the
quiescent state in the frequency range up to $\Delta_{\textrm{Al}}/h \approx
50$\,GHz, i.e. within the working frequency range of the Josephson qubits of
different types.

Finally, the modelled behavior of the simple RSFQ networks (JTL and TFF) is
qualitatively similar to that of the conventional (resistively shunted)
circuits. A preliminary evaluation of the ranges of functionality was found to
be quite good. The simulation of more complex circuits can be easily performed
using the slightly modified code PSCAN developed earlier for the tunnel
junction model \cite{Odintsov,PSCAN}. So, RSFQ networks with SIN-shunts
implemented in the shell that surrounds a Josephson qubit core offer a
promising approach to achieve joint RSFQ-qubit operation.

\section{Acknowledgments}

We wish to thank D.\,V.~Balashov, F.~Chiarello, S.\,V.~Lotkhov, F.~Maibaum and
M.~Wulf for useful discussions and assistance at different stages of this work.
The work was supported by the EU through the RSFQubit and SQUBIT-2 projects and
DFG through the project Ni\,253/7-1.



\begin{references}
\bibitem{comment} The condition $\Delta_{\textrm{SIS}} \gg \Delta$ can be
fulfilled, for example, for the pair of a Nb SIS junction and a SIN junction
with Al superconductor electrode.
\end{references}
\end{document}